\documentclass[%
 reprint,
 amsmath,amssymb,
 aps,
]{revtex4-1}

\usepackage{graphicx}
\usepackage{dcolumn}
\usepackage{color}
\usepackage{xspace}
\usepackage{units}

\def\mathclap#1{\text{\hbox to 0pt{\hss$\mathsurround=0pt#1$\hss}}} 
\newcommand{\msub}[1]{_{\text{#1}}}

\DeclareFontFamily{U}{euc}{}
\DeclareFontShape{U}{euc}{m}{n}{<-6>eurm5<6-8>eurm7<8->eurm10}{}%
\DeclareSymbolFont{AMSc}{U}{euc}{m}{n} 
\DeclareMathSymbol{\umu}{\mathord}{AMSc}{"16}

\begin{document}
\preprint{APS/123-QED}

\title{A low energy muon spin rotation and point contact tunneling study of niobium films prepared for superconducting cavities}

\author{Tobias Junginger} 
\email{Tobias.Junginger@helmholtz-berlin.de}
\affiliation{Helmholtz-Zentrum Berlin fuer Materialien und Energie (HZB), Germany}
\altaffiliation{TRIUMF Canada's National Laboratory for Particle and Nuclear Physics, Vancouver}
\author{S. Calatroni}
\affiliation{European Organisation for Nuclear Research (Cern),Geneva Switzerland}
\author{A. Sublet}
\affiliation{European Organisation for Nuclear Research (Cern),Geneva Switzerland}
\author{T. Prokscha}
\affiliation{Paul Scherrer Institut (PSI), Villigen, Switzerland} 
\author{T. Proslier}
\affiliation{Argonne National Laboratory, USA}
\affiliation{Commissariat de l'\'energie atomique et aux \'energies renouvelables, France}
\author{Z. Salman}
\affiliation{Paul Scherrer Institut (PSI), Villigen, Switzerland} 
\author{A. Suter} 
\affiliation{Paul Scherrer Institut (PSI), Villigen, Switzerland} 
\author{G. Terenziani}
\affiliation{European Organisation for Nuclear Research (Cern),Geneva Switzerland}
\altaffiliation{Sheffield University, UK}
\author{J. Zasadzinski}
\affiliation{Illinois Institute of Technology, Chicago, USA}

\date{\today}

\begin{abstract}
Point contact tunneling (PCT) and low energy muon spin rotation (LE-$\mu$SR) are used to probe, on the same samples, the surface superconducting properties of micrometer thick niobium films deposited onto copper substrates using different sputtering techniques: diode, dc magnetron (dcMS) and HIPIMS. The combined results are compared to radio-frequency tests performances of RF cavities made with the same processes. Degraded surface superconducting properties are found to correlate to lower quality factors and stronger Q slope. In addition, both techniques find evidence for surface paramagnetism on all samples and particularly on Nb films prepared by HIPIMS.

\end{abstract}

\maketitle

\section{Current limitations of niobium on copper cavities}

Superconducting cavities prepared by coating a micrometer thick niobium film on a copper substrate enable a lower surface resistance compared to bulk niobium at \unit[4.5]{K} the operation temperature of several accelerators using this technology, like the LHC or the HIE-Isolde at CERN. Additionally Nb/Cu cavities are more cost-effective and do not require magnetic shielding. Thermal stability is enhanced avoiding quenches \cite{Calatroni200695}. 

Despite these advantages, this technology is currently not being considered for accelerators requiring highest accelerating gradient $E\msub{acc}$ or lowest surface resistance $R\msub{S}$ at temperatures of \unit[2]{K} and below. The reason for this is that $R\msub{S}$ increases strongly with $E\msub{acc}$. The origin of this field dependent surface resistance has been the subject of many past and recent studies \cite{Darriulat,CalatronSRF2001,CalatroniSRF03,Junginger_PRSTAB_2015}, but is still far from being fully understood. The film thickness of about \unit[1.5]{$\mu$m} is large compared to the London penetration depth $\lambda\msub{L}$ of about \unit[32]{nm}. Differences in the performance (apart from thermal conductivity issues \cite{palmieri2015thermal}) should therefore be correlated to the manufacturing procedure and the resulting surface structure. Since no single dominant source can be expected, several hypotheses need to be addressed individually to identify their origin and possibly mitigate their influence on the surface impedance and SRF cavity performances.
\\

\subsection{Surface magnetism, a possible source of RF dissipation}

A possible source of dissipation under RF fields is surface magnetism in the oxide layer; magnetic impurities have a well known deleterious effect on superconductivity, causing inelastic scattering of the Cooper pairs \cite{Abrikosov1957199} that increase the surface resistance $R\msub{S}$ and lower the quality factor $Q\propto$ 1/$R\msub{S}$ of superconducting RF cavities. The presence of localized magnetic moments on bulk niobium samples was revealed for the first time by Casalbuoni et al. in 2005 \cite{casalbuoni2005surface} by SQUID magnetometry. The AC magnetic susceptibility was measured in an external magnetic field of \unit[0.7]{T} on Nb samples prepared with standard SRF cavity treatments, i.e. buffered chemical polishing (BCP), electropolishing (EP) and low temperature baking at about 120 $^{\circ}$C. In all cases, the samples displayed a Curie-Weiss behavior below \unit[50]{K} indicative of the presence of localized magnetic moments with antiferromagnetic correlation. 

Later, Proslier et al. \cite{Proslier_APL_2008} used the point contact tunneling spectroscopy (PCTS) technique to study samples prepared in a similar manner to SRF cavities. These samples exhibit a broadened density of states (DOS) compared to ideal superconductors with a finite zero bias conductance attributed to inelastic scattering processes. In addition, some tunnel junctions show large zero bias conductance peaks (ZBP) that survive in an external magnetic field above the second critical field H$_{C2}$ of Nb ($\approx$\unit[0.24]{T}). The temperature dependence of the ZBP measured under an external DC magnetic field of \unit[0.55]{T} \cite{Proslier_2011} was found to be consistent with Kondo tunneling which describes the interaction between tunneling electrons and localized magnetic moments present in the oxides or near the interface native oxide/bulk niobium. In light of these results, the broadened DOS was further examined and the tunneling conductance spectra were fitted using the Shiba theory that calculates the effects of diluted magnetic impurities on the superconducting DOS. The very good agreement between the theory and the experimental data provides a plausible microscopic origin for the inelastic scattering processes found in Nb samples.

Further theoretical work was carried out to quantify the influence of magnetic impurities on the RF surface impedance \cite{kharitonov2012surface}. The numerical simulations showed that magnetic impurities, present in small quantities ($\sim$ few hundred ppm), cause the saturation of the surface impedance at low temperature, an effect called the 'residual resistance' which is observed experimentally on all cavities. 

More recently, PCT studies have been extended to cut-outs from cavities that showed pronounced medium field Q-slope. Some samples, labeled as 'hot spots', were cut out from regions that showed strong dissipation as measured by thermometry \cite{Proslier_SRF2013} and compared to samples cut out from regions that did not, labeled as 'cold pots'. The results show that tunneling spectra measured on hot spots have in average higher inelastic scattering parameters and lower gap values than cold spots. In addition, about \unit[30]{$\%$} of the junctions measured in hot spot samples showed zero bias peaks whereas almost none were measured in cold spot samples. These results further indicate a correlation between the presence of magnetic impurities and the dissipation measured in SRF cavities.

Various explanations have been proposed to account for the presence of magnetic moments at surfaces. In general, several metal oxide layers and oxide nanoparticles \cite{Sundaresan} were found to develop magnetic properties in otherwise nonmagnetic oxides. It is interesting to note that only oxide nanoparticles with a size of about \unit[10]{nm} showed magnetic moments whereas if the material was pressed into a bar and sintered the samples became diamagnetic. Besides, native niobium oxides are thermodynamically stable with substantial off-stoichiometry and Cava et al. \cite{cava1991electrical} showed that oxygen deficient Nb$_2$O$_{5-\delta}$ samples develop magnetic moments associated with unpaired Nb4d electrons. Recent density functional theory (DFT) simulations \cite{DFTNb2O5} revealed that these local moments form 1d spin chains with weak antiferromagnetic correlations coupled to perpendicularly oriented conducting nano-wires that can be traced back to the NbO$_6$ octahedral structure. 
 
It has to be noted that although magnetic impurities can cause surface dissipation, other impurity phases with lower superconducting properties than niobium such as Nb hydrides \cite{trenikhina2015nanostructural} or carbon contamination \cite{ZasadzinskiCarbon} that have been found on cavity-grade niobium samples and cut outs can also be responsible for higher surface resistance. For Nb on Cu cavities, another dissipation mechanism extrinsic to superconductivity, the variable Nb-Cu interface thermal conductivity caused by inhomogeneous Nb film adhesion to the Cu substrate, has been proposed \cite{palmieri2015thermal,calatroni2016performance} to explain the strong Q-slope measured. 




\section{Sample preparation and surface characterization}


Several coating techniques are presently exploited at CERN. Direct Current  Magnetron Sputtering (dcMS) is the technique of choice for the 4-cell \unit[352]{MHz} elliptical cavities previously used for the Large Electron-Positron (LEP) collider \cite{benvenuti1991superconducting} and the single cell \unit[400]{MHz} elliptical cavities currently installed in the LHC \cite{Chiaveri99}. This technique is constantly being developed by making use of smaller scale test cavities operating at 1.3 or \unit[1.5]{GHz}, which allow a faster turn-over and easier implementation of new coating and surface processing ideas. A dummy cavity is used for the preparation of the coupons, located at its equator, i.e. the cavity region with the largest diameter corresponding to the peak magnetic field. Typical dcMS sputtering conditions \cite{Darriulat} require a DC power of \unit[1]{kW} in Kr plasma, resulting in \unit[1.5]{$\mu$m} thick films in \unit[15]{minutes} coating time. Coatings are performed at \unit[150]{$^\circ$C}. 

High-power Impulse Magnetron Sputtering (HIPIMS) is also being developed on \unit[1.3]{GHz} elliptical cavities as a potential technique to mitigate field dependent losses \cite{TerenzianiSRF13}. In HIPIMS the high current, high density Kr plasma ionises the Nb from the target thus allowing more dense films to be formed. The HIPIMS coating was performed on a coupon with the same identical setup as for dcMS, at similar average power of about \unit[1]{kW}. The peak plasma current was \unit[200]{A}, with \unit[200]{$\mu$s} pulses at a repetition rate of about \unit[100]{Hz}. The coating duration, owing to the lower efficiency of HIPIMS, is adapted in order to obtain a similar thickness as for dcMS.

\begin{figure*}[t]
	\centering
	\includegraphics[width=\textwidth]{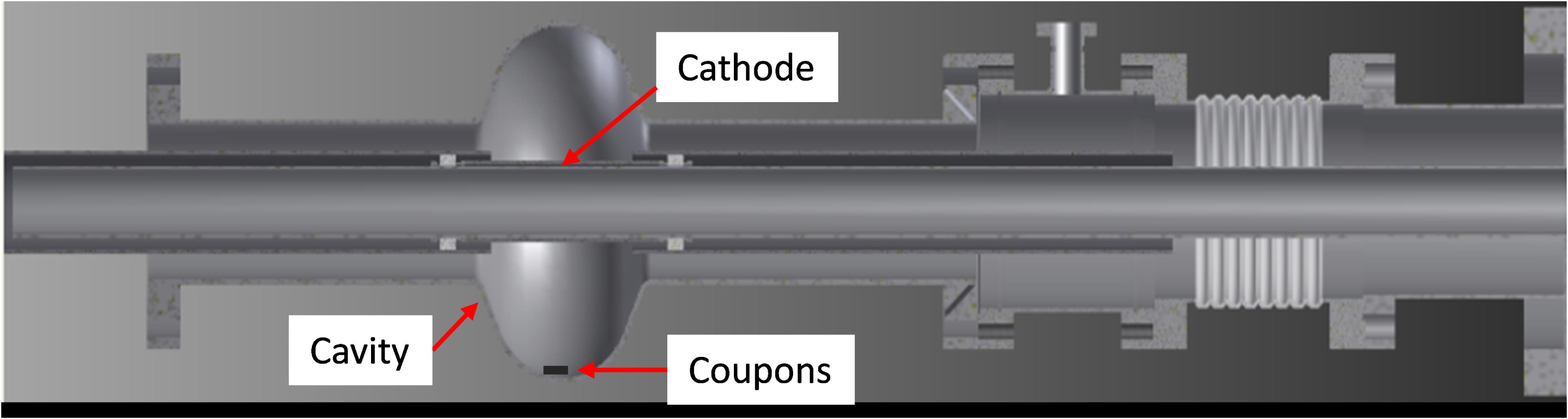}
	\includegraphics[width=\textwidth]{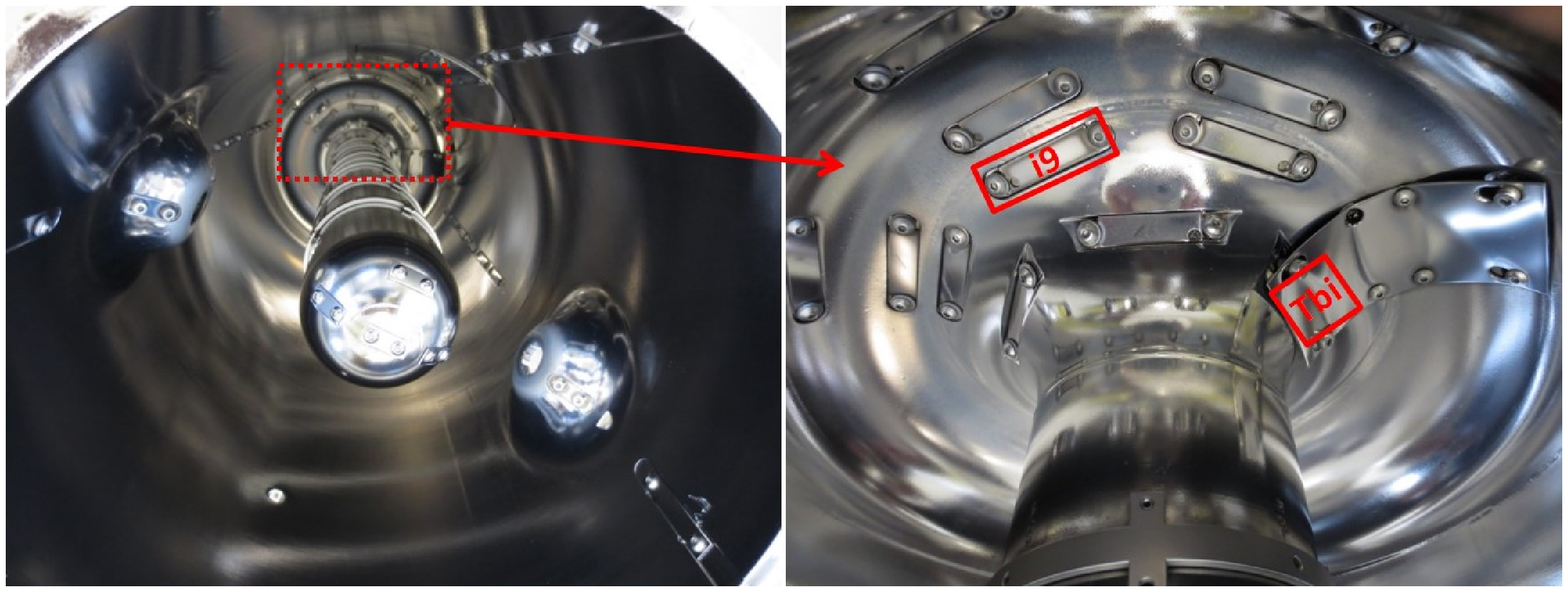}
	\caption{Top: Schematics of the \unit[1.3]{GHz} sputtering setup as used to produce the dcMS and HIPIMS samples. The samples are located at the equator, the region of largest diameter. Bottom: Position of the HIE-ISOLDE Nb/Cu samples i9 and Tbi taken from the mock-up cavity for PCT and LE-$\mu$SR measurements respectively.}
	\label{fig:Sample_positions}
\end{figure*}

For the HIE-Isolde quarter wave \unit[100]{MHz} cavities DC bias diode sputtering is used which is easier to apply to these more complex geometries \cite{sublet2015developments}. A mock-up cavity has been designed as a sample holder to characterize the Nb layer as described in \cite{bartova2015characterization}. The samples measured in this study are taken from the top of the cavity, named i9 and TBi for for PCT and LE-$\mu$SR measurements respectively, as depicted in Fig.\,\ref{fig:Sample_positions}. This position corresponds to the peak RF magnetic field region of the cavity, which is one of the most critical parts form RF point of view. The thickness of the niobium layer at these position is \unit[2.7]{$\mu$m}.
The coating has been made in 15 successive runs of 23 minutes each at high temperature (from \unit[300]{$^\circ$C} to \unit[620]{$^\circ$C}) with \unit[5.5]{hours} of cool-down in between each run resulting in a total time of 4 days. The setup consists of a cylindrical Nb cathode coaxially mounted around the inner conductor of the cavity and is used to coat inner and outer conductor. The sputtering is made at \unit[8]{kW} with Ar.

\begin{figure}[tbh]
	\centering
	\includegraphics[width=0.9\columnwidth]{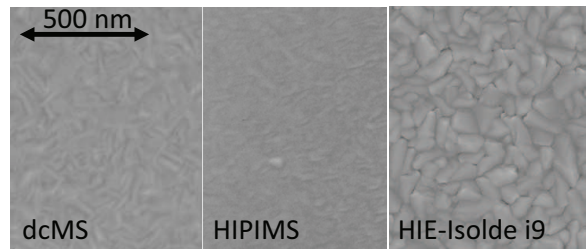}
	\caption{Scanning electron microscope (SEM) images of the three samples used for the point contact tunneling measurements.}
	\label{fig:SEM}
\end{figure}

Different microstructures are obtained for the three sputtering techniques. In particular the grain size of the HIE-Isolde dc bias coated samples ($\approx$\unit[500-1000]{nm}) \cite{bartova2015characterization} is roughly 10 times larger compared to HIPIMS \unit[50-100]{nm} and dcMS \unit[100-200]{nm} \cite{TerenzianiSRF13}, see Fig.\,\ref{fig:SEM} In this study one sample of each technique has been investigated by PCT and LE-$\mu$SR. 

\section{Experimental results}

In this section results from point contact tunneling performed at Argonne National Laboratory and low energy muon spin roation (LE-$\mu$SR) carried out at the Paul Scherrer Institute are presented. 
The samples tested by these two methods were produced simultaneously at CERN. The results are being correlated to the RF performance of cavities produced using the same sputtering setups and parameters. 

\subsection{RF cavity tests}

RF performance tests have been carried out on cavities prepared by all three coating techniques using the phase locked loop technique \cite{Padamsee:1116813}. In these tests one usually obtains the unloaded quality factor $Q_0$ as a function of the accelerating gradient $E\msub{acc}$. Both parameters do not only depend on the material properties but also on the cavity geometry. To compare test results from differently shaped cavities in terms of their material properties it is more meaningful to plot the surface resistance $R\msub{S}$ as a function of the peak surface magnetic field $B\msub{p}$ instead of $Q_0$ versus $E\msub{acc}$. While $E\msub{acc}$ and $B\msub{p}$ are directly proportional to each other, $R\msub{S}$ can be obtained from $Q_0$ using the geometry factor $G$, which has to be calculated by a simulation for realistic cavity shapes \cite{Padamsee:1116813}. Assuming $R\msub{S}$ does not vary over the cavity surface it can be written as
\begin{equation}
R_{S}=G/Q_0.
\end{equation}       
Table \ref{tab:tRFParameters} lists the geometry factor and the ratio $B\msub{p}/E\msub{acc}$ for all cavities used in this study. Note that the dcMS and HIPIMS coatings have been deposited on cavities of same shape but slightly different size. Therefore $G$ and $B\msub{p}/E\msub{acc}$ are identical and only $f$ differs. 
\begin{table}[b]
	\centering
		\begin{tabular}{l|c|c|c}
		Coating & $f$[GHz]  & $B\msub{p}/E\msub{acc}$[mT/mV] & G[$\Omega$]\\
		\hline
		HIPIMS & 1.3 & 4.26 & 270 \\
		dcMS & 1.5 & 4.26 & 270 \\
		HIE-Isolde &	0.1 & 9.5 & 30.8 \\
		ISAC-II & 0.106 & 10 & 19.1 \\
		\end{tabular}
	\caption{Relevant RF parameters of the test cavities taken from literature; HIPIMS/dcMS\cite{PhysRevSTAB.3.092001}, HIE-Isolde\cite{zhang2015frequency}, ISAC-II\cite{facco2001superconducting}.}
	\label{tab:tRFParameters}
\end{table}

To compare the RF performance of the DC biased diode coating deposited on the \unit[100]{MHz} HIE-Isolde cavities to the dcMS and HIPIMS coatings, not only the cavity shape, but also the different RF frequency needs to be considered. Losses from thermally activated quasiparticles, so called 'BCS losses' scale exponentially with temperature and quadratically with frequency. A practical formula to quantify these is \cite{Padamsee:1116813}
\begin{equation}
R\msub{BCS}=A\omega^2 \exp\left({-\frac{\Delta(0)}{k\msub{B}T}}\right),
\label{eq:RBCS}
\end{equation}
where $\omega=2\pi f$, $\Delta(0)$ is the superconducting energy gap at \unit[0]{K} and $A$ depends on material parameters. At a temperature of \unit[4.5]{K} $R\msub{BCS}$ equals only a few n$\Omega$ at $f$=\unit[100]{MHz}, while at \unit[1.3]{GHz} it is at least several \unit[100]{n$\Omega$} in case of niobium. Operation of superconducting cavities at such high $R\msub{S}$ becomes uneconomical. Therefore the elliptical HIPIMS and dcMS cavities are designed for and usually measured at \unit[1.8]{K}, while the operation temperature of the HIE-Isolde accelerator is \unit[4.2]{K}. However test results of the best HIPIMS cavity at \unit[4.2]{K} are also available. 

Additional to the BCS losses there are residual losses, which do not depend on temperature. The total surface resistance $R\msub{S}$ thus equals
\begin{equation}
R\msub{S}=R\msub{BCS}+R\msub{res}.
\end{equation}
The origin of the residual resistance $R\msub{res}$ and how it scales with frequency is not completely understood. Experimentally it has been found that both $R\msub{BCS}$ and $R\msub{res}$ depend on the applied RF magnetic field strength. Especially Nb/Cu cavities show a strong increase of $R\msub{res}$ with applied RF field \cite{Darriulat,SergioResidual}. 

All cavity substrates were prepared using a polishing agent named SUBU, which is a mixture of sulfamic acid, hydrogen peroxide, $n$-butanol and ammonium citrate which is a standard technique for Nb/Cu cavities \cite{Darriulat}. The elliptical cavities also received electropolishing (EP) prior to SUBU, which can provide a smoother substrate. RF test results from the following cavities have been chosen for this study:
\begin{itemize}
	\item{HIPIMS M.2.3: Best performing HIPIMS cavity with EP+SUBU substrate (tested at 1.8 and \unit[4.2]{K}),}
	\item{dcMS H.6.8: Best performing dcMS cavity with EP+SUBU substrate (tested at \unit[1.8]{K} only),}
	\item{HIE-Isolde QP.1.4: Best performing HIE-Isolde cavity with SUBU substrate (tested at \unit[4.2]{K} only).}
\end{itemize}

\begin{figure}[htb]
	\centering
	\includegraphics[width=0.9\columnwidth]{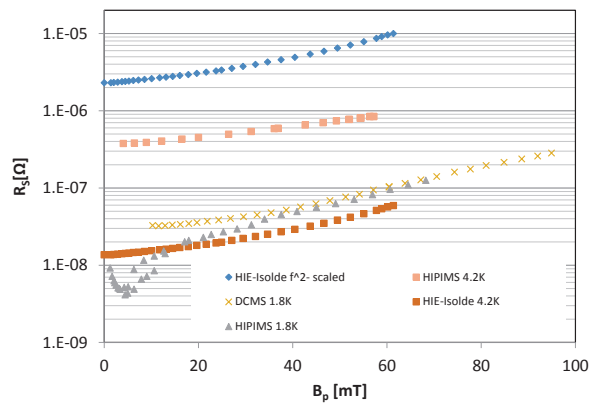}
	\caption{Surface resistance $R\msub{S}$ as a function of peak surface magnetic field $B\msub{p}$ for three thin film cavities.}
	\label{fig:RvsB_films}
\end{figure}

Figure \ref{fig:RvsB_films} shows $R\msub{S}$ as a function of $B\msub{p}$ for these three cavities. The performance of the two elliptical cavities at \unit[1.8]{K} is fairly similar. The HIE-Isolde cavity has the lowest $R\msub{S}$ at \unit[60]{mT} even if this result is compared to the measurements at \unit[1.8]{K} of the elliptical cavities. However if $R\msub{S}$ is quadratically scaled to \unit[1.3]{GHz} as suggested by Eq. \ref{eq:RBCS} the performance is far weaker compared to the HIPIMS cavity at same temperature. This plot illustrates the difficulty to compare the performance of the elliptical HIPIMS and dcMS cavities to the HIE-Isolde quarter wave cavity due to different frequency and geometry. An heuristic approach to work around this issue is to compare each cavity type to similar ones made of bulk niobium and then in turn compare the performance differences to each other to get at least a rough idea about the RF performance independent of resonant frequency and cavity shape.         
%
%
Suitable devices for this comparison are the elliptical \unit[1.3]{GHz} cavities and \unit[100]{MHz} quarter wave resonators from TRIUMF, both manufactured from niobium sheets. 
In particular RF test results from these two cavities are taken:
\begin{itemize}
	\item{PAV-2: A single-cell \unit[1.3]{GHz} elliptical cavity produced as a prototype cavity for the ARIEL linac and tested at 1.8 and \unit[4.2]{K}. The RF parameters are identical to the HIPIMS cavity.}
	\item{ISAC-II-SCB5-3: One of the best performing \unit[106]{MHz} cavities used in the ISAC-II accelerator. The RF parameters are listed in Tab. \ref{tab:tRFParameters}}
\end{itemize}

\begin{figure}[htb]
	\centering
	\includegraphics[width=0.9\columnwidth]{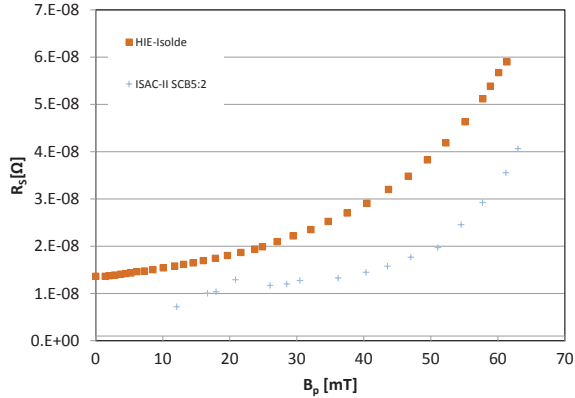}
	\caption{Surface resistance $R\msub{S}$ as a function of peak surface magnetic field $B\msub{p}$ for two quarter wave cavities at \unit[4.2]{K}.}
	\label{fig:Isac_Isolde}
\end{figure}

\begin{figure}[htb]
	\centering
	\includegraphics[width=0.9\columnwidth]{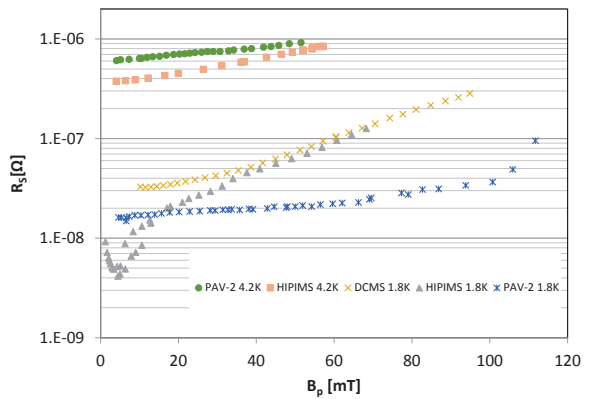}
	\caption{Surface resistance $R\msub{S}$ as a function of peak surface magnetic field $B\msub{p}$ for three elliptical cavities.}
	\label{fig:Ellipticals_Rs}
\end{figure}

Figure \ref{fig:Isac_Isolde} shows $R\msub{S}$ for the two quarter wave cavities at \unit[4.2]{K}. The residual resistance $R\msub{res}$ at low $B\msub{p}$ is significantly larger for the HIE-Isolde cavity while the increase of $R\msub{S}$ with field is similar. Figure \ref{fig:Ellipticals_Rs} shows $R\msub{S}$ for the three elliptical cavities. At \unit[4.2]{K} the performance of the HIPIMS cavity exceeds the bulk niobium cavity especially at low field. This can be related to the shorter electron mean free path, resulting in a lowered $R\msub{BCS}$ \cite{Darriulat}. At \unit[1.8]{K} and low field $R\msub{S}$ is lowest for the HIPIMS cavity. However, at this temperature the well-known limitation of the niobium on copper technology becomes apparent; $R\msub{res}$ increases strongly with the applied field strength. One might argue that the performance of the HIE-Isolde coating is weaker compared to the other two coatings, since it does not outperform bulk niobium at \unit[4.2]{K}. However, it has to be noted that the performance of the \unit[1.3]{GHz} cavities at \unit[4.5]{K} is clearly dominated by BCS losses while in case of the \unit[100]{MHz} quarter wave resonators residual losses are still relevant. Measurements at lower temperature could serve to quantify the extend of the two contributions. Unfortunately, such data is not available for the resonators being studied here.    

\subsection{Point contact tunneling (PCT)}

\begin{figure*}[t]
	\centering
	\includegraphics[width=\textwidth]{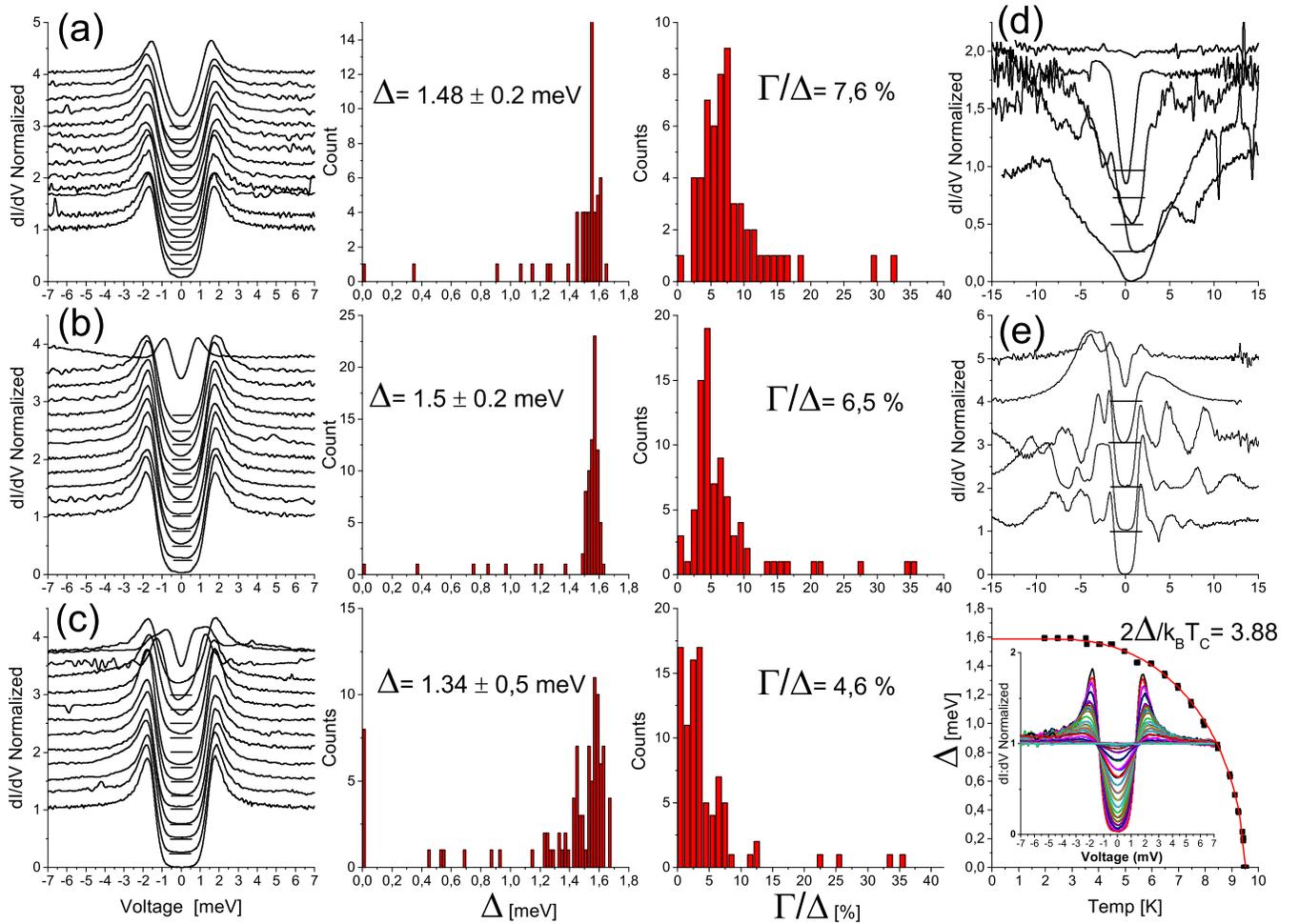}
	\caption{Summary of the PCTS measurement performed on 3 niobium samples deposited on Cu substrate with 3 different deposition methods; (a) HIPIMS, (b) standard DC magnetron sputtering or dcMS and (c) diode sputtering. For each sample, 80 to 100 junctions were measured. A set of about 10 representative normalized tunneling conductance curves are represented on the left (shifted by 0.25 for more clarity). The statistics of the superconducting gap $\Delta$ and inelastic parameter $\Gamma/\Delta$ as extracted from the fits are displayed in the middle. The temperature dependence of a characteristic tunnel junction is shown on the right with the corresponding temperature dependence of $\Delta$. (d) Example of conductance spectrum shifted by 0.25 that show no hint of superconductivity and considered as normal metal junctions $\Delta$ = 0 meV. (e) Example of tunnel junction showing very peculiar background for V$\ge\Delta$, the curves have been shifted by 1 for more clarity.}
	\label{fig:PCT}
\end{figure*}

PCT is a powerful technique that measures the surface density of states (DOS) and the related superconducting properties: the superconducting gap $\Delta$, the critical temperature $T\msub{c}$ and the phenomenological inelastic scattering parameter $\Gamma$ using the native oxide layer as tunnel barrier \cite{groll2015point}. These parameters play a crucial role in the SRF cavity performances; at low external magnetic field intensities (neglecting vortices dynamic) $R\msub{S}$ depends on the DOS at the fermi level $\nu (E_f)$ that in turn depends on $\Delta$, $\Gamma$ and the temperature for a superconductor. Qualitatively, in an ideal superconductor and for temperatures $T \ll T\msub{c}$, $\nu (E_f) \propto e^{-\Delta/k_B.T}$ and lower $\Delta$ values increase the $R\msub{S}$ and therefore lower the quality factor $Q_0$. Inelastic scattering processes however yield a minimal $\nu (E_f) \sim \Gamma/\Delta$ value independent of temperature that entails a saturation of the surface resistance at low temperature, $R\msub{res}$, and limits the cavity quality factor. The total surface resistance, $R\msub{S}$, is a combination of these two effects.

In order to understand the SRF cavity performances it is therefore crucial to measure the surface DOS for each new surface treatment or deposition technique. We present here the results obtained by PCT on three Nb on Cu films few microns thick and prepared with the different deposition techniques introduced above. 

The summary of the point contact tunneling results on the three samples measured at \unit[1.6]{K} is displayed in Fig.\ref{fig:PCT}. The gap, $\Delta$, and the phenomenological inelastic scattering parameter, $\Gamma$ values are extracted from fits of the conductance curves using the Blonder-Tinkham-Klapwijk (BTK) model \cite{Dynes,Blonder}. For each sample, the mean values of the $\Delta$ and $\Gamma/\Delta$ distributions are shown. As apparent in Fig.\,\ref{fig:PCT}, the HIPIMS (a) and the dcMS samples (b) have similar mean $\Delta$ (1.48 meV and 1.5 meV) and $\Gamma/\Delta$ (7.6$\%$ and 6.5$\%$) values. These data suggest that the surface impedance of unprocessed Nb film on copper cavities made by HIPIMS and dcMS should be the same, in agreement with RF tests done on cavities as seen in Fig.\,\ref{fig:Ellipticals_Rs}. Furthermore, these results are very similar to previous measurements done on bulk niobium cut out from cavities treated with standard recipes and consistently, very close $R\msub{res}$ values of $\sim$\unit[10]{n$\Omega$} at low field have been measured on both types of cavities at low RF field.

The Nb/Cu sample made by diode sputtering Fig. \ref{fig:PCT}(c) shows a much broader gap distribution than the other two samples with a very low average $\Delta$ of \unit[1.3]{meV} and a $\Gamma/\Delta$ value of 4.6$\%$. These results affect the $R\msub{S}$ in opposite way: lower average $\Delta$ should increase $R\msub{S}$ whereas lower $\Gamma/\Delta$ tends to decrease it. In the absence of RF tests done on cavities with the same frequency and temperature it is hard to correlate the PCT data obtained on the diode sputtered sample to the RF performances.
%
%
All the samples show a majority of the tunneling spectra, between $\sim$ 70 $\%$ to $\sim$ 90 $\%$ depending on the samples, with high quality superconducting DOS characteristic of bulk Nb. However, the data also systematically reveals very low gap values $\leq$ 1.3 meV and sometimes even non-superconducting junctions ($\Delta$=0 meV) such as the one displayed in Fig. \ref{fig:PCT}(d). Investigating these tunneling spectrum in more detail, we found that for all the samples the low $\Delta$ junctions have large $\Gamma/\Delta$ values of $\geq$ 7 $\%$ up to 35 $\%$ indicative of some weak superconducting and potentially highly dissipative spots present on the surface. These spectra account for 4 to 8$\%$ of the junctions measured. 

Similar results have only been obtained previously on hot spots bulk niobium samples cut out from cavities with a strong medium field $Q$-slope \cite{Proslier_SRF2013}. Such correlation between strong $Q$-slope and very weak superconducting properties is not surprising; as the external magnetic field intensity is increased in the cavity, regions of the surface with lower gap values than bulk Nb (caused by proximity effect, magnetic or some non-magnetic impurities, deleterious phases such as hydrides or carbides) will serve as easy flux entry points and decrease the depairing critical current value. These effects combined suggest a runaway mechanism and weakened superconductivity generates dissipation exponentially $dR\msub{S}/dH \propto \alpha R\msub{S}$, causing a medium field Q slope observed experimentally on some bulk Nb cavities and on all the Nb on Cu cavities (Fig. \ref{fig:RvsB_films}). What is more surprising is the fact that these regions do not seem to influence the value of the $Q$ factor at very low field; bulk niobium cavities that do not have strong $Q$-slope or/and very low gap values as measured by PCT, show similar Q values at very low field $\sim 2.10^{10}$ corresponding to a residual resistance of less than \unit[10]{n$\Omega$}. A possible explanation is that the good thermal conductivity of the surrounding clean Nb can remove efficiently the small amount of power generated by these spots at very low field.

The microscopic origin for these degraded properties is unknown yet, there are however major structural and chemical differences between Nb films grown on Cu and bulk Nb samples: the average grain size $\approx$ \unit[50-1000]{nm} is 2 to 3 orders of magnitude smaller than for fine grain bulk niobium and the impurity concentration \cite{Bloess} (mostly C and O) is about 10-100 times larger in Nb films. It is therefore possible that the low $\Delta$ junctions corresponds to grain boundaries regions with weakened superconductivity due to impurity segregation or porosities near the surface as seen on some thin Nb film cross sections \cite{calatroni2016performance}. Assuming a circular grain shape of size $d$ with a weak superconducting ring of thickness 2$\xi$ around the grain boundary, the probability, $P$ of probing such a region with a constant steps size between junction measurement is $\propto 2\xi/d$. Assuming $\xi$=\unit[40]{nm} ($\ll d$) and d$\sim$ \unit[1000]{nm}, $P\sim$ 8$\%$. In this picture, the majority of the tunnel junctions with high quality superconducting DOS comparable to bulk niobium cavities are measured within the grains with gap values at T=\unit[1.6]{K} consistent with bulk Nb of 1.5 to \unit[1.6]{meV} and a typical BCS temperature dependence as shown on Fig. \ref{fig:PCT}(c) with critical temperatures $T\msub{c}$ from 9.25 to \unit[9.4]{K}. While this simple model is consistent with the data measured on dc diode sputtered sample (10$\%$), it fails to explain why the degraded properties are more often observed for this sample compared to the other two with smaller grain size, suggesting that for the dcMs and the HIPIMS samples grain boundaries do not systematically host low $T\msub{C}$ phases. 

Some tunnel spectra show very strong peaks in the conductance curves at a voltage $\ge \Delta$, see Fig. \ref{fig:PCT}(e) and (d). These features are present in all samples and particularly on the diode sputtered one. The peak positions are, in general, not symmetric with respect to the Fermi level and can hardly be attributed to charging effects. They are can be attributed however to localized defects in the tunnel barrier, here the Nb native oxide, with well defined energy levels within the oxide band gap. Coincidentally, the same sample also has the lowest mean average gap. It is therefore possible that these conductance peaks originate from impurities sometimes present within the tunnel barrier, sometimes underneath the oxide in the superconductor where they would be responsible for the lower gap values. To confirm this hypothesis more structural and chemical analysis are needed. These results have been reproduced on another diode sputtered sample made under similar conditions (not shown).

\begin{figure}[t]
	\centering
	\includegraphics[width=1\columnwidth]{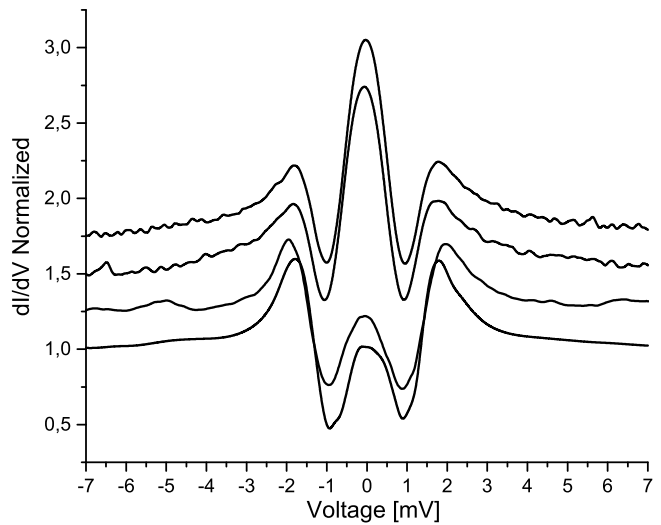}
	\caption{(a) Example of conductance curves displaying the zero bias conductance peak obtained on the HIPIMS sample, the curves are shifted by 0.25 for more clarity.}
	\label{fig:Kondo}
\end{figure}

Finally, some tunneling spectra such as the ones displayed in Fig. \ref{fig:Kondo}, show strong zero bias conductance peaks (ZBP) identical to spectra measured previously \cite{Proslier_2011} on bulk Nb samples cut out from cavities with strong medium field Q-Slope and characteristic of the Kondo effect. Most zero bias peaks are found on the HIPIMS sample with $\sim 12\%$ of the junctions measured, the HIE-Isolde sample has only a few $\sim 2\%$, and the dcMS sample has $\sim 5\%$. Assuming these ZBPs are caused by magnetic impurities located along grain boundaries, the probability of probing such a feature should be proportional to the grain size, which is consistent with the trend found in the measurements. These results confirm the presence of magnetic impurities on the surface of these niobium on copper samples, probably localized at grain boundaries, with the highest concentration on the HIPIMS one. Similar results have been reproduced on another HIPIMS sample (not shown). The latter sample also has the highest mean inelastic scattering parameter. Fitting the quasiparticle peaks symmetric with respect to the Fermi level one can estimate $\Gamma/\Delta$ and $\Delta$ values for each spectrum that shows a ZBP; the average $\Gamma/\Delta$ value is 12 $\%$ for the HIPIMS and 9.5 $\%$ for the dcMS sample, well above the overall average for both samples. This result indicates that indeed the presence of magnetic impurities tends to increase the average inelastic scattering values and hence the dissipation in cavities. The HIPIMS and dcMS cavities performances however are qualitatively similar (low field $R\msub{S}$ and Q-slope) indicating that magnetic impurities do not play a dominant role in dissipation processes for thin Nb films.

\subsection{LE-$\mu$SR measurements}
\label{LE-muSR measurements} 
Muon spin rotation relaxation and resonance ($\mu$SR) is an experimental technique providing information at an atomic level on the 
physical properties of matter. It is based on the implantation of spin-polarized muons and the measurement of the evolution of their spin with time due to the magnetic field experienced by the particle \cite{yaouanc2011muon}. The main applications of this technique are in magnetic materials and in superconductors. Muon beams of different energies and corresponding implantation depth are available at PSI, TRIUMF, J-PARC, ISIS and RIKEN-RAL. For the studies of the niobium surfaces we are interested in the outermost \unit[100]{nm} of the sample. A suitable beam is currently only available at the Laboratory for Muon Spin Spectroscopy (LMU) at PSI \cite{prokscha2008new}. 

\subsubsection{Direct measurement of the London penetration}

\begin{figure}[t]
   \centering
	 \includegraphics[width=0.95\columnwidth]{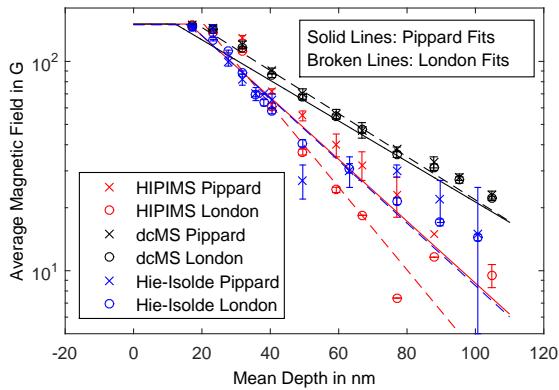}
   \caption{Penetration of the magnetic field in three Nb/Cu samples at $T\approx$\unit[3.5]{K}}
	\label{fig:Penetration}
\end{figure}

Low energy muon spin rotation (LE-$\mu$SR) enables the measurement of the magnetic field inside a sample as a function of depth. An external magnetic field with a value below $H\msub{c}$ (type-I superconductor) or $H\msub{c1}$ (type-II superconductor) can be applied parallel to the sample surface to probe the London penetration depth. This has been done first in 2000 \cite{Jackson2000} and more recently for bulk niobium samples prepared for superconducting cavities \cite{Romanenko_muSR}. In the latter study Romanenko et al. showed that there is a depth dependent mean free path for samples exhibiting low losses under RF fields after a \unit[120]{$^{\circ}$C} bakeout. These results gave the motivation to study niobium on copper samples to search for similar variations in the penetration profile.
%



Figure \ref{fig:Penetration} shows the penetration profiles obtained from analyzing the LE-$\mu$SR data using the musrfit software \cite{Suter}. Two fit models were used to calculate each data point. A simple Gaussian model based on London theory and a numerical time-domain model based on the non-local Pippard/BCS model \cite{Suter2005}. The global Pippard fits have been performed to the whole raw data set. For the HIE-Isolde sample this fit did not converge. Therefore no global Pippard fit is displayed in Fig. \ref{fig:Penetration} for this sample. The deposition here is done in several runs with breaks in between. This might have resulted in an inhomogeneous penetration profile inconsistent with one set of global fit parameters. The penetration depth $\lambda$ of the dcMS sample is so large that non-local effects become unimportant. The obtained $\lambda$ is therefore consistent for the local and the non-local fit. For the HIPIMS sample, however, non-local effects are more relevant. Clear deviations from the results between local and non-local data analysis were found. 

Using \unit[1.3]{GHz} superconducting cavities the penetration depth was also obtained by measuring the frequency shift as a function of temperature for the dcMS and the HIPIMS technique, see Appendix and \cite{Junginger_PRSTAB_2015}. These values are systematically higher than the values obtained by $\mu$SR, see Tab. \ref{tab:Penetration}. It should be noted that with RF one only measures the penetration depth change in the temperature range between about 4 and \unit[9]{K} and obtains the penetration depth at \unit[0]{K}, $\lambda_0$, as a fit parameter. Also this method relies on an exponential decay of the magnetic field. 
\begin{table}[bht]
	\caption{Penetration depth value in nm obtained by $\mu$SR and RF. $\mu$SR results are for local/non-local fit. Error range for RF includes several cavity measurements.}
	\centering
		\begin{tabular}{|l|c|c|c|}
		\hline
	&	dcMS	& HIPIMS & Hie-Isolde\\
	\hline
$\mu$SR	& 43(5)/45(2)	&  22(6)/29(1) & 29(5)/-\\
\hline
RF	& 60(10)	& 47(2) & - \\
		\hline	
\end{tabular}
	\label{tab:Penetration}
\end{table}
%


Comparing RF and LE-$\mu$SR results from the different samples we find that with HIPIMS and diode sputtering at high temperature films with a lower penetration depth and therefore lower impurity concentration can be produced. However the dcMS cavity, corresponding to the sample with the largest $\lambda$, performs best under RF fields of the three thin film cavities tested, see Fig.\,\ref{fig:RvsB_films}. The cause of the field dependent residual surface resistance must be related rather to localized impurities, possibly along grain boundaries or to other loss mechanism like the adhesion of the film on the substrate \cite{palmieri2015thermal}.

\subsubsection{Zero field measurements} 
\begin{figure}[tbp]
   \centering
	 \includegraphics[width=0.8\columnwidth]{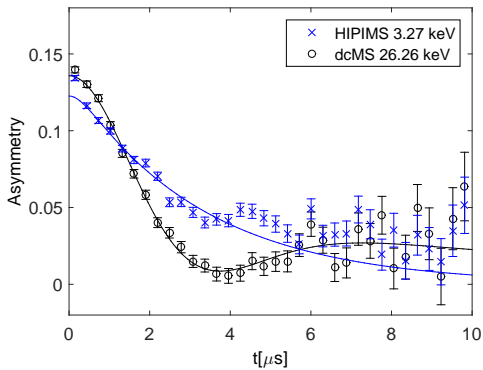}
   \caption{Asymmetry functions obtained at zero field and $T\approx$\unit[3.5]{K} for the HIPIMS sample at the surface and the dcMS sample at about \unit[100]{nm} depth. The HIPIMS sample shows strong signs of muon dynamics possibly related to surface magnetism. The asymmetry function of the dcMS sample recovers to about 1/3 of its initial value, which signifies that each muon experiences a static magnetic field, while the fields seen by different muons are randomly distributed. Note: the asymmetry function is proportional to the spin polarization where a proportionality factor is given by the spectrometer design.}
	\label{fig:ZF}
\end{figure}

A strength of $\mu$SR compared to e.g. NMR is the possibility to measure internal magnetic fields under zero applied field conditions. Here the implanted muon is measuring the local magnetic field at the muon site with high sensitivity. In case of a static random field distribution, e.g. nuclear dipole fields, and the absence of muon diffusion, the muon spin polarization is given by

\begin{equation}\label{eq:statGKT}
 P_{\rm ZF}^{\rm stat.}(t) = \frac{1}{3} + \frac{2}{3} \left[ 1 - (\sigma t)^2 \right] \exp\left[-\frac{1}{2} (\sigma t)^2\right],
\end{equation}

\noindent where $\sigma / \gamma_\mu$ is the second moment of the magnetic field distribution at the muon site, and $\gamma_\mu$ is the muon gyromagnetic ratio. A point dipole calculation for the expected interstitial muon site in Nb results in $\sigma\msub{mag} \simeq 0.5\, \mathrm{\mu s}^{-1}$. If other static and randomly distributed sources of local magnetism, e.g. magnetic impurities, are present the total $\sigma$ becomes  
\begin{equation}
\sigma^2 = \sigma\msub{dip}^2 + \sigma\msub{mag}^2,
\end{equation}
where $\sigma\msub{dip}$ is the nuclear dipole contribution and $\sigma\msub{mag}$ is originating of other sources, e.g. magnetic impurities. 


The static polarization function $P_{\rm ZF}^{\rm stat.}(t)$ (Eq.\,(\ref{eq:statGKT})) is modified in presence of fluctuating magnetic fields and/or muon diffusion. In a strong collision model the resulting muon spin polarization function is

\begin{eqnarray}\label{eq:strong-collision-model}
  P(t) &=& P_{\rm ZF}^{\rm stat.}(t) \exp(-\nu t) +  \\ \nonumber 
       & & +\, \nu \int_0^t dt^\prime \left\{ P(t-t^\prime) P_{\rm ZF}^{\rm stat.}(t^\prime) \exp(-\nu t^\prime) \right\},
\end{eqnarray}

\noindent where $\nu$ is the fluctuation rate ($1/\nu = \tau$ being the correlation time). In the so called extreme motional narrowing limit $\nu \gg \sigma$, Eq.(\ref{eq:strong-collision-model}) applied to Eq.(\ref{eq:statGKT}) results in

\begin{equation}\label{eq:motional-narrowing-limit}
 P_{\rm ZF}^{\rm fast}(t) = \exp\left(- \lambda_{\rm mn} t\right),\qquad \lambda_{\rm mn} = \frac{2 \sigma^2}{\nu}.
\end{equation}

Fast diffusion as well as fast fluctuating (nano to femto-second range) magneitc impurities will therefore yield an exponential depolarization function and in the extreme motional narrowing limit a constant depolarization function. For dilute impurities this function will be superimposed with the depolarization function of muons not sensing areas with magnetic impurities.



Zero field measurements in order to search for hints of magnetic impurities were carried out at energies of 3.3 and \unit[26.3]{keV} corresponding to mean muon stopping depths $d$ of 17.3 and \unit[104.8]{nm} for all three samples.
Individual curves have been fitted to the dynamic Gaussian Kubo-Toyabe function, Eq.\,\ref{eq:strong-collision-model}. 
The dcMS and the HIE-Isolde sample show only weak signs of dynamics and broadening beyond the expected nuclear dipole broadening, while the HIPIMS sample shows clear signs at both measured energies (see Tab. \ref{tab:ZF}). 

For the HIPIMS sample $\sigma$ is significantly exceeding the nuclear dipole contribution, \footnote{For comparison measurements on bulk niobium samples of different treatments (EP, EP+\unit[120]{$^{\circ}$C}, BCP, N doping) gave very similar $\nu$ values of around \unit[0.4-0.48]{Mhz} for all samples \cite{Romanenko_pc}.} and the largest $\nu$ is found. This sample has the smallest grain size. Hopping should thus be strongly diminished \cite{luetkens2003diffusion}, indicating a different origin for the observed muon dynamics. Comparison with the PCT results suggests that magnetic impurities located at grain boundaries might be the cause.  


\begin{table}[bth]
	\caption{Fluctuation rate $\nu$ and static width $\sigma$ of three Nb/Cu samples for different temperatures $T$ and mean muon stopping depths $d$ obtained from zero field measurements.}
	\centering
		\begin{tabular}{l|c|c|c|c}
	Sample & $d$ & $T$  & $\nu$  & $\sigma$ \\
				 & [nm] & [K]&   [MHz] & [MHz] \\
	\hline
	HIPIMS & 17.3  & 3.75 & $3.0\pm2.2$ & $0.65\pm0.16$\\	
				&  104.8 & 3.83 & $4.2\pm2.1$ & $0.71\pm0.27$ \\	
				\hline
 	dcMS & 17.3  & 3.30 & $0.43\pm0.05$ & $0.49\pm0.03$\\
				&  104.8 & 3.55 & $0.11\pm0.03$ & $0.50\pm0.012$ \\
				\hline
 	Hie-& 17.3  & 2.70 & $0.34\pm0.04$ & $0.51\pm0.011$ \\
	Isolde		& 100.0  & 2.66 & $0.28\pm0.03$ & $0.429\pm0.008$\\ 
				
	
\end{tabular}
	\label{tab:ZF}
\end{table}

\subsubsection{Longitudinal field studies of the HIPIMS sample}
The zero field measurements presented above indicate that surface magnetism is indeed present at least for the HIPIMS sample. 
However muon diffusion as the cause for these fluctuations cannot be ruled out. Also it has to be noted that the depolarization rate, $\sigma$, and the fluctuation rate, $\nu$, are strongly correlated if both are used as individual fit parameters. 

\begin{figure}[htb]
   \centering
	 \includegraphics[width=0.8\columnwidth]{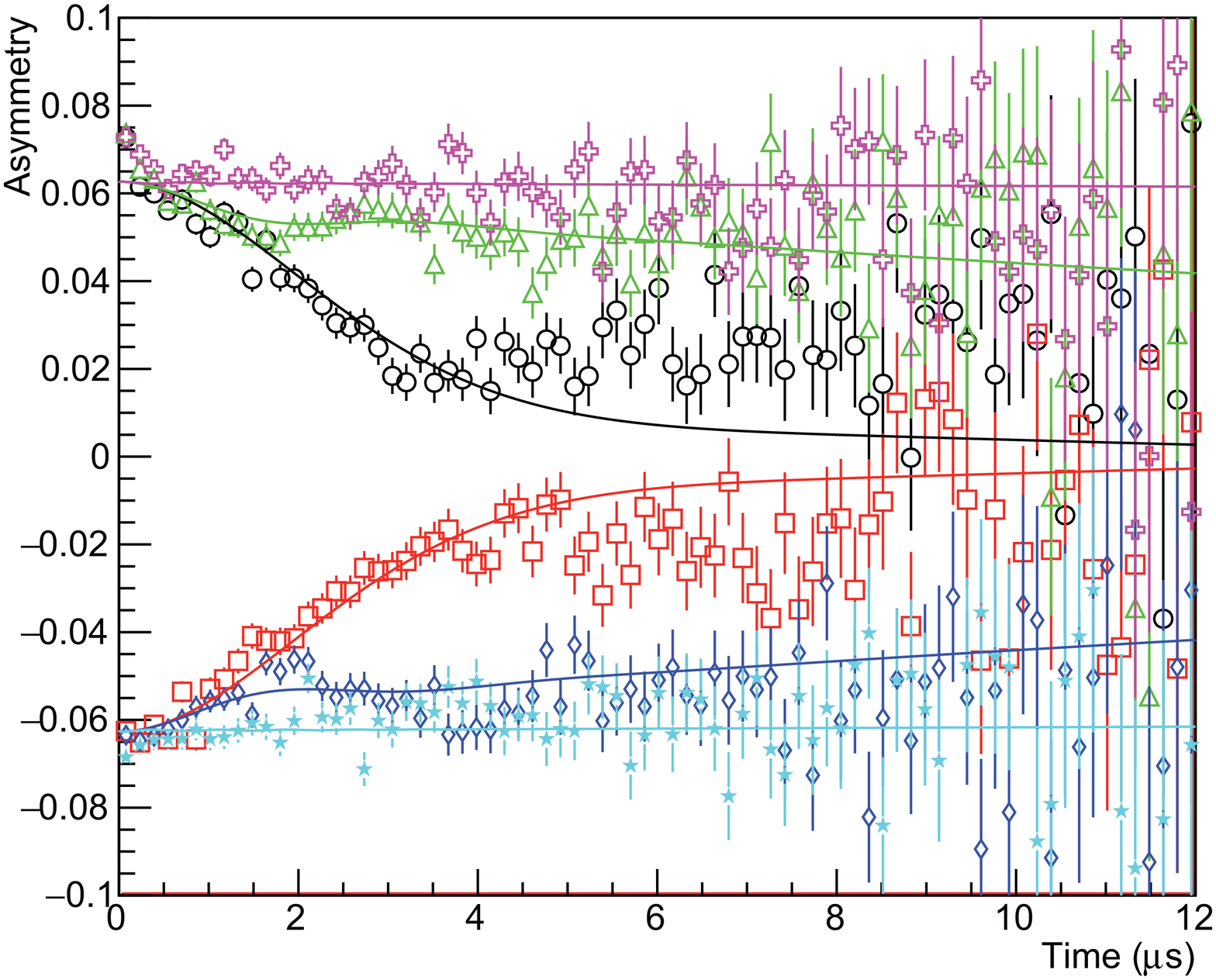}
   \caption{Asymmetry function versus time at \unit[50]{K} for the HIPIMS sample in a longitudinal field of 0,2 and 10 mT. }
	\label{fig:50K}
\end{figure}

Of the samples investigated above the HIPIMS sample showed the strongest hints of magnetic impurities. It was therefore decided to study the very same sample as investigated by PCT above in depth with LE-$\mu$SR. \footnote{Here it should be noted that the two studies were first started in parallel with different samples prepared simultaneously in the same deposition run. After completion of the PCT measurements these samples became available for the LE-$\mu$SR study presented in the following.}  
By applying different longitudinal fields and fitting the data simultaneously to a common $\sigma$ value it is possible to get a better estimate for $\sigma$ and $\nu$. In order to further suppress possible systematic errors half of the data has been taken with the muon spin aligned in the direction of muon momentum and half of the data with the spin in opposite direction. Figure\,\ref{fig:50K} displays the asymmetry function versus time at \unit[50]{K} for longitudinal fields of 0, 2 and \unit[10]{mT}. The upper curves display data where the initial polarization was in the direction of the muon momentum. 

\begin{figure}[tbh]
   \centering
	 \includegraphics[width=0.8\columnwidth]{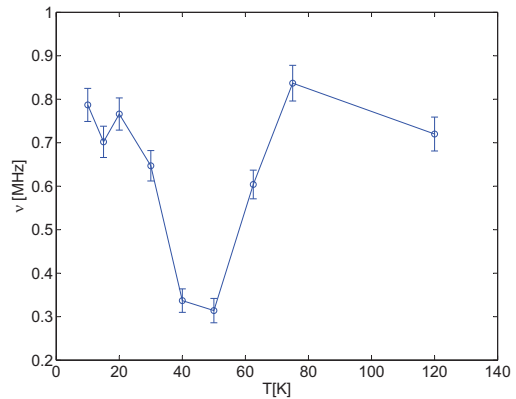}
   \caption{Hop rate $\nu$ as a function of temperature for the HIPIMS sample. The line is a guide to the eye.}
	\label{fig:HopvsT}
\end{figure}

The hop rate,$\nu$, as a function of temperature is displayed in Fig.\,\ref{fig:HopvsT}. To understand the results they need to be compared with previous $\mu$SR studies on niobium. For a brief literature review refer to the Appendix. The most apparent feature is the minimum observed at \unit[50]{K}. In studies where the muon has been implanted in the bulk this feature has also been observed and was correlated to interstitial impurities \cite{hartmann1983trap}. A more recent study on cavity grade bulk niobium \cite{Grassellino2013} has also found a minimum at \unit[50]{K}. It has to be noted that in this study the variation of the hopping rate with temperature was much stronger than what is observed here. Also in \cite{Grassellino2013} $\nu$ was obtained from zero field measurements only. Samples prepared by sputter coating have impurity concentrations about 10-100 times larger than RRR niobium sheets. The total impurity concentration of the samples analyzed here can be estimated to be in the \unit[1000]{ppm} range \cite{Bloess}. A sample with such a high interstitial impurity concentration has been analyzed in \cite{niinikoski1979muon}. The depolarization rate found was similar as for material of higher purity at low temperature in accordance with the expectation for static muons. 
Unlike for samples of higher purity however $\sigma$ and therefore $\nu$ did only weakly depend on temperature indicating stronger trapping \footnote{Note that in this study $\sigma$ was taken as the only parameter to account for muon dynamics.}. 

The temperature dependence displayed in Fig.\,\ref{fig:HopvsT} significantly differs from what has been observed in \cite{niinikoski1979muon} and therefore suggests an additional cause for the muon dynamics in addition to hopping, which might be indeed paramagnetic impurities as suggested from the zero field LE-$\mu$SR and the PCT studies.

\subsubsection{Zero field studies of the HIPIMS sample with a N$_2$ overlayer}

\begin{figure}[tbh]
   \centering
	 \includegraphics[width=0.8\columnwidth]{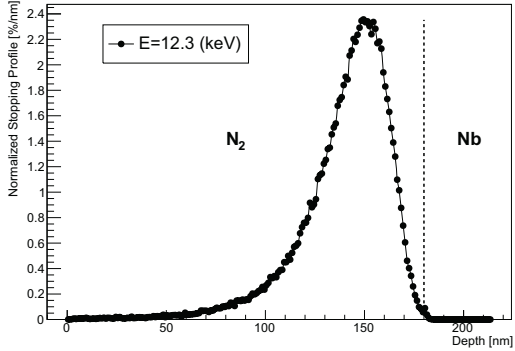}
   \caption{Muon stopping profile for the applied bias voltage of \unit[12.3]{keV}.}
	\label{fig:180nm_N2}
\end{figure}

The following study is carried out to distinguish between magnetic and non-magnetic impurities. If these impurities are magnetic they would contribute to the depolarization in two ways. First they would, like nonmagnetic impurities, act as trapping sites. This effect would decrease the hopping rate $\nu$. Second if the paramagnetic correlation time would be in the femto to nano-second range it would result in an increased $\nu$, since the muon would see a time varying field in its lifetime. Unfortunately it is impossible to extract the influence of the latter mechanism from the data. The correlation time of paramagnetic moments in the niobium oxides is also unknown. To overcome this limitation a \unit[180(20)]{nm} N$_2$ layer has been grown on top of the niobium sample and the muons have been stopped in this N$_2$ layer, close to the niobium. Figure\,\ref{fig:180nm_N2} displays the muon stopping profile obtained from the Monte-Carlo code \texttt{TRIM.SP} \footnote{This program has been developed at MPI Garching by W.\,Eckstein \cite{biersack1984sputtering, ecksteincomputer} and adopted and experimentally verified for muons \cite{morenzoni2002implantation}.}.  

\begin{figure}[tbh]
   \centering
	 \includegraphics[width=0.8\columnwidth]{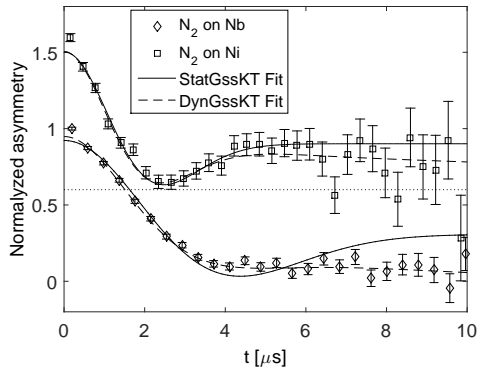}
   \caption{Normalized asymmetry function of muons stopped in an N$_2$-overlayer on top of the HIPIMS sample and a Ni plate at \unit[10]{K}. The latter data set has been shifted along the y-axis by 0.6.}
	\label{fig:Noverlayer}
\end{figure}

From previous studies it is known that muons are static, that means they do not diffuse, in nitrogen as grown under the given conditions. Nevertheless an additional measurement has been performed to verify this important assumption. 
%
%
A \unit[1.83]{$\mu$m} thick nitrogen layer has been grown on a Ni coated sample plate. The Ni coating has a thickness of about \unit[2]{$\mu$m}. The N$_2$ layer thickness
has been chosen since the static stray fields of the Ni plate increase the depolarization rate $\sigma$ and therefore enhance the sensitivity in this cross-check experiment. The asymmetry function has been measured at \unit[10]{K} in zero field, see Fig.\,\ref{fig:Noverlayer}. The data can be well described by a dynamic Kubo-Toyabe depolarization function with a very low hop rate or even a static Gaussian Kubo-Toyabe depolarization function, see Tab.\,\ref{tab:Noverlayer}. This clearly demonstrates that the muons are indeed static, i.e. not diffusing, in the N$_2$ layer. 
For the muons stopped in N$_2$ on top of the HIPIMS sample the static Gaussian Kubo-Tuyabe function cannot give a reasonable fit, since the asymmetry function does not relax to $1/3$ of its initial value as expected for static muons. In fact the signal clearly shows a dynamic response, which further supports the presence of magnetic impurities in these films, since muon diffusion is fully suppressed here and hence the origin of the fluctuation rate can only be caused by magnetic fluctuations present in the niobium. Using a dynamic Kubo-Toyabe function instead gives an excellent fit to the data, see Fig.\,\ref{fig:Noverlayer} and Tab.\,\ref{tab:Noverlayer}.
\begin{table}[bth]
	\caption{Fluctuation rate $\nu$ and static width $\sigma$ obtained from zero field measurements for muons stopped in N$_2$ on top of a HIPIMS Nb sample and a Ni plate at \unit[10]{K}. Values are for static/dynamic Gaussian Kubo-Tuyabe function.}
	\centering
		\begin{tabular}{l|c|c|c|}
	       & $\nu$  & $\sigma$ \\
			 & [MHz] &  [MHz] \\
	\hline 
	N$_2$ on HIPIMS	&-/0.29(5) & 0.394(6)/0.470(15) \\
	N$_2$ on Ni &-/0.077(35) & 0.699(19)/0.734(25) \\

\end{tabular}
	\label{tab:Noverlayer}
\end{table}
%
%

\section{Summary and Conclusion}
In this study niobium films for SRF application, sputter coated by three different techniques on copper substrates, have been investigated by point contact tunneling (PCT) and low energy muon spin rotation and relaxation (LE-$\mu$SR). The results from these material science techniques have been compared to RF measurements (surface resistance and penetration depth) of cavities prepared with the same setups and sputtering parameters. The goal was to get insight into the field dependent residual surface resistance currently limiting the application space of this technology to moderate accelerating gradients.

For one sample (HIPIMS) the combined results give strong evidence for magnetic impurities, while there is little evidence found on the other two samples prepared by DC diode and DC magnetron sputtering (dcMS). The fraction of tunneling junctions exhibiting zero bias peaks, indicative of magnetic impurities, correlates inversely to the grain size. It is therefore likely that these impurities are located along grain boundaries, since the smaller the grain size the more likely it is to probe a grain boundary with PCT. Comparing this finding to the RF performance, it is however unlikely that pairbreaking caused by these magnetic impurities plays the dominant role for the field dependent residual surface resistance, since the dcMS and the HIPIMS cavities show a very similar RF performance. A comparison to the HIE-Isolde quarter wave resonator with the dc diode coating is difficult due to the lower resonance frequency and the unknown frequency dependence of the residual resistance. 

The dc diode sputtered sample showed smaller gap values $\Delta$ than the other two samples. Assuming that all RF losses scale quadratically with frequency, like BCS losses, this is in agreement with its weaker RF performance. Very low $\Delta$ values are found on all samples indicative of weak superconducting spots. These areas account for about 4-8\% of all measured junctions. For bulk niobium such a feature has only been found for samples cut out from cavities with strong field dependent losses. It has been shown in \cite{Dhakal} that such sub gaps states can be suppressed by high temperature baking. However, for Nb/Cu cavities such a treatment is not possible due to the lower melting temperature of copper. 

RF frequency shift and LE-$\mu$SR results show that with dc bias coating at higher temperatures and HIPIMS one can obtain a film with significantly longer electron mean free path and therefore lower impurity concentration compared to standard dcMS. However, neither of the two techniques yields an RF performance exceeding dcMS films, indicating that the electron mean free path does not play an important role for the field dependent residual surface resistance of Nb/Cu cavities either.


The samples used in this study have been prepared in the same deposition chambers used for the actual cavities. However, it has to be noted that the RF losses have not been measured on the very same samples. It is well known that RF losses in superconducting cavities may not be evenly distributed but are often dominated by a few hot spots on the surface. Future work should then concentrate in testing hot and cold spots cut out from Nb/Cu cavities or samples measured in an RF sample test cavity by PCT and LE-$\mu$SR to directly correlate the RF performance to the superconducting properties of Nb/Cu samples. 



\section{Acknowledgment}
The $\mu$SR measurements were performed at the Swiss Muon Source (S$\mu$S), at the Paul Scherrer Institute in Villigen, Switzerland. This work has been funded partly by a Marie Curie International Outgoing Fellowship and the EuCARD-2 project of the European Community’s 7th Programme and by the U.S. Department of Energy, Office of Sciences, Office of High Energy Physics, early Career Award FWP 50335 and Office of Science, Office of Basic Energy Sciences under Contract No. DE-AC02-06CH11357. We thank P. Kolb for providing the RF data of the PAV-2 cavity and Ignacio Aviles Santillana for providing the SEM images.

\section{Appendix I: Stopping profiles}
\begin{figure}[tbh]
   \centering
	 \includegraphics[width=0.8\columnwidth]{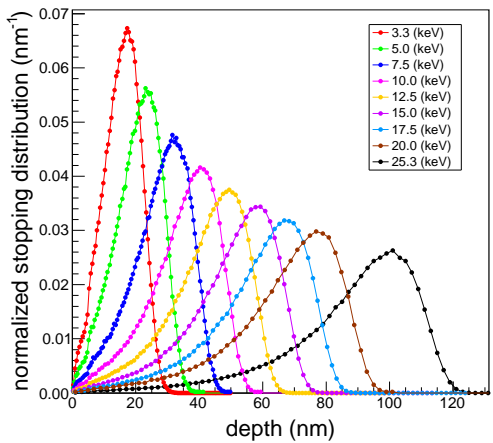}
   \caption{Stopping profiles of muons in niobium of different energy calculated with the TRIM.SP software.}
	\label{fig:Nb-muon-stopping-distribution}
\end{figure}

Figure \ref{fig:Nb-muon-stopping-distribution} displays the muon stopping profiles obtained from the Monte-Carlo code \texttt{TRIM.SP}. This program has been developed at MPI Garching by W.\,Eckstein \cite{biersack1984sputtering, ecksteincomputer}.
%
%
and adopted and experimentally verified for muons \cite{morenzoni2002implantation}.

\section{Appendix II: RF penetration depth measurements}

\begin{figure}[t]
   \centering
	 \includegraphics[width=0.8\columnwidth]{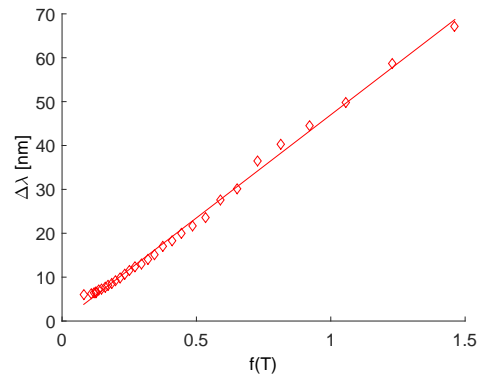}
   \caption{Penetration depth change measured on a \unit[1.3]{GHz} HIPIMS cavity as a function of $f(T)=1/\sqrt{1-(\frac{T}{T\msub{c}})^4}-1$ \cite{Junginger_PRSTAB_2015}.}
	\label{fig:RFpendepth}
\end{figure}

Figure \ref{fig:RFpendepth} displays the penetration depth change obtained from the RF frequency shift for a HIPIMS cavity \cite{Junginger_PRSTAB_2015}. The penetration depth at \unit[0]{K}, $\lambda_0$ is extracted from a least squares fit to the Gorter-Casimir expression \cite{Gorter1934306}
\begin{equation}
\lambda(T)=\frac{\lambda_0}{\sqrt{1-(\frac{T}{T\msub{c}})^4}},
\end{equation} 
with the critical temperature $T\msub{c}$ as a second independent fit parameter.
The penetration depth, $\lambda$, is directly related to the electron mean free path $l$, a measure of the impurity concentration, via \cite{1953}
\begin{equation}
\lambda(l)=\lambda(l\rightarrow \infty)\sqrt{1+\frac{\pi\xi_0}{2l}},
\end{equation}
where the London penetration depth for infinite mean free path $\lambda(l\rightarrow \infty)$ and the BCS coherence length, $\xi_0$, are material parameters for which literature values are available. In case of bulk niobium $\lambda(l\rightarrow \infty)$=\unit[32]{nm} and $\xi_0$=\unit[39]{nm} have been reported \cite{MattisBardeenTheory}, while LE-$\mu$SR results for rather clean films find $\lambda_L$=\unit[27(3)]{nm} \cite{Suter2005}. For dcMS sputter coated niobium films Benvenuti et al. derived $\lambda$=\unit[29(3)]{nm} and $\xi_0$=\unit[33]{nm} from RF measurements \cite{Darriulat}.

\section{Appendix III: Brief literature review of the temperature dependence of the hopping rate for niobium of different purity }

The depolarization of positve muons implanted in niobium has been extensively studied since the late 1970s \cite{birnbaum1978anomalous,niinikoski1979muon,brown1979mu+,boekema1982diffusion,hartmann1983trap,hartmann1984initial}. In these publications muon diffusion and trapping is studied for samples covering a wide range in purity containing interstitial and substitutional impurities. While earliest studies where carried out in transverse field configuration it was in 1982 that Boekema et al. \cite{boekema1982diffusion} pointed out that zero field measurements can unambiguously distinguish between static and dynamic muons, since the static muon will have its polarization value recover to 1/3 of its initial value for long times (see Eq.(\ref{eq:statGKT})). It has to be noted that previous studies usually analyze their data in terms of a Gaussian or an exponential depolarization rate $\lambda$, only and not in terms of depolarization, $\sigma$, and fluctuation/hopping rate, $\nu$, using the dynamic Gaussian Kubo-Toyabe depolarization function as given in Eq.(\ref{eq:strong-collision-model}) and first derived in Ref.\,\cite{Hayano} in the context of Lapalce transfoms. In the former $\lambda$ accounts for the muon dynamics. A low $\lambda$ corresponds to a high hopping rate in the corresponding two parameter fit (see Eq.(\ref{eq:motional-narrowing-limit})). This can be understood in terms of the concept of motional narrowing \cite{yaouanc2011muon}, i.e. in case of fast diffusion different muons will in average sense almost identical fields during their lifetime leading to a small $\lambda$. 

For niobium with an impurity concentration of less than \unit[1]{ppm} it has been found that $\nu$ does not depend on temperature \cite{hartmann1983trap}. In this case a depolarization rate significantly smaller than expected from the second moment of the nuclear dipole field is found indicating fast diffusion. Generally impurities give a more complex behavior of the depolarization or hopping rate as function of temperature. For moderately pure niobium with an impurity concentration of several \unit[100]{ppm} of interstitial impurities a two plateau structure is generally observed and explained within a two trap model. At a temperature below about \unit[16]{K} the muon rapidly finds an extended shallow trap and remains there during the observation time. As the temperature is increased the muon becomes thermally detrapped. If the temperature is further increased the muon is mobile enough to find deeper more localized traps. This yields a second minimum of the hopping rate around \unit[60]{K}. 

It has to be noted that all these studies have been carried out with so called surface muons, implanted about \unit[0.3]{mm} deep in the bulk of the samples. The studies presented here however use a low energy muon beam of variable energy implanting the muons in the nanometer range, therefore not the bulk properties but the oxide layer, the \unit[100]{nm} outermost metal surface and the oxide/metal interface are probed. It is important to note that it is not directly possible to reveal the origin of muon dynamics. The technique only shows whether the muons sense a constant or a time varying magnetic field during their lifetime. Usually, in good metals, the time varying signal is interpreted as the muon diffusing in the material. However paramagnetic impurities with a correlation time on the order of a femto to nano-seconds also yield a time varying field. 

\end{document}